\documentclass[showpacs,preprintnumbers,amsmath,amssymb,floatfix]{revtex4}

\usepackage{epsfig,psfrag}
\usepackage{dcolumn}
\usepackage{bm}
\usepackage{graphicx}
\usepackage{color}

\begin{document}

\title{Josephson current through a quantum dot with spin-orbit 
coupling}
\author{L. Dell'Anna,$^{1}$ A. Zazunov,$^{2}$ R. Egger,$^1$
 and T. Martin$^{3,4}$}    
\affiliation{
${}^1$~Institut f\"ur Theoretische Physik, Heinrich-Heine-Universit\"at,
D-40225  D\"usseldorf, Germany \\
${}^2$~LPMMC CNRS, 25 av. des Martyrs, F-38042 Grenoble, France\\
${}^3$~Centre de Physique Th{\'e}orique, Case 907 Luminy,
F-13288 Marseille, France\\
${}^4$~Universit{\'e} de la M{\'e}dit{\'e}rann{\'e}e, 
F-13288 Marseille, France }
\date{\today}

\begin{abstract}
The equilibrium Josephson current through a nanoscale
multi-level quantum dot with Rashba or Dresselhaus 
spin-orbit coupling $\alpha$ has been computed. 
The critical current can be drastically modified
already by moderate $\alpha$. In the presence of 
a Zeeman field, Datta-Das-like oscillatory
dependencies on $\alpha$ are predicted.
\end{abstract}

\maketitle

\section{Introduction}

The Josephson effect provides a fundamental signature
of phase coherent transport through mesoscopic samples
with time reversal symmetry \cite{jj}.
Here, we reconsider the theory of the Josephson effect through
a quantum dot (QD) in a two-dimensional electron gas (2DEG), 
taking into account Rashba and/or Dresselhaus spin orbit (SO) couplings
\cite{rashba,egger}.  Despite the large recent interest concerning
SO effects in QDs, mainly caused by quantum
information \cite{dot} and spintronics applications \cite{datta}, 
the question of how the Josephson current is modified by
SO couplings has barely been addressed.
Naively, since SO couplings do not break
time reversal symmetry, one may expect that they will not
affect the Josephson effect at all.  This expectation seems
basically confirmed by the existing theoretical studies:
In Ref.~\cite{chalmers}, the Josephson current 
through a {\sl perfectly contacted}\ 2DEG with Rashba
coupling is predicted to only show SO-related modifications in the 
simultaneous presence of a Zeeman field.  In Ref.~\cite{krive},
the same conclusion has been reached for a {\sl single-channel}\ 
conductor with arbitrary contact resistance.  Moreover, 
in Ref.~\cite{feigelman}, by a generalization
of earlier scattering approaches \cite{furusaki,beenakker},
the case of a wide 2DEG with arbitrary
contact resistance was studied, where the semiclassically averaged
Josephson current is again found to be insensitive to Rashba SO
couplings. However, the presence of a Josephson current through the mesoscopic 
system necessarily breaks time reversal invariance and therefore 
we reexamine this expectation in the present paper.  

As 2DEG devices based on InAs-related materials are known
to exhibit strong gate-tunable \cite{nitta} Rashba SO couplings, 
and possibly Dresselhaus SO couplings, we study this
problem here from the perspective of
multi-level QD physics.  
Supercurrents through related devices have been probed by several
experiments; for recent work, see 
Refs.~\cite{takayanagi,schaepers,giazotto,ebel,bouchiat}.
It is thus not only of academic interest to quantitatively examine 
the effects of Rashba/Dresselhaus SO couplings on the 
equilibrium Josephson current.
Moreover, very recently, gate-tunable supercurrents
through thin InAs nanowires have been reported
 \cite{doh,vandam}, revealing complex 
current-phase relations such as $\pi$-junction behavior. 
Although we study a 2DEG geometry, our results are also 
relevant for such nanowires:
the transport channels reside in a surface charge layer, and
Rashba terms due to narrow-gap and strong confinement
fields dominate over all other SO couplings. In those experiments, 
Josephson currents through few-level dots have already been achieved.

In the present work, we arrive at the surprising conclusion
that {\sl SO couplings have huge effects on the Josephson
current in nanoscale  multi-level dots}. Depending on the parameter
regime, the critical current can be largely suppressed or 
enhanced.  In a single-level dot, SO effects on the Josephson
current vanish unless
there is also a Zeeman field, but in multi-level dots no such
restriction applies.  The predicted strong dependence on SO couplings 
should be observable in state-of-the-art experiments.
We also show that {\sl oscillations in the critical
current}\ $I_c$ as a function of the distance $L$ between the 
lead contacts appear under suitable conditions, 
resembling Datta-Das \cite{datta} spin precession effects. 

The structure of the remainder of this paper is as follows.  In
Sec.~\ref{Sec.II}, the model is described.
The Josephson current is calculated in Sec.~\ref{Sec.III},
and results are presented in Sec.~\ref{Sec.IV} for the simplest
cases of one and two spin-degenerate levels. We conclude in 
Sec.~\ref{Sec.V}. Some details concerning the derivation of the
Josephson current have been delegated to an Appendix.  

\section{Model}
\label{Sec.II}

We study a QD formed by a confinement potential 
$V({\bf r})$ within a 2DEG, ${\bf r}=(x,y)$.  
The total Hamiltonian reads: 
\begin{equation}
H=H_D+H_T+H_L+H_R
\end{equation}
 and using the QD fermion creation operator 
$d^\dagger_\sigma({\bf r})$ for spin 
 $\sigma=\pm$, the isolated QD is described by (we put $\hbar=k_B=1$,
and spin summations are often left implicit)
\begin{equation} \label{hd}
H_D= \int d{\bf r} 
\ d^\dagger({\bf r}) \left( \frac{(-i\nabla + {\bf a})^2 -2\alpha^2 }{2m}
+ {\bf b}\cdot \vec\sigma + V\right) d({\bf r}),
\end{equation} 
where $d = \left( d_\uparrow, d_\downarrow \right)^T$,
$m$ is the effective mass, ${\bf b}=(b_x,b_y,b_z)$ is 
a constant external Zeeman field (including gyromagnetic and Bohr
magneton factors), and $\vec \sigma=(\sigma_x,\sigma_y,\sigma_z$) with  
standard Pauli matrices.  
Orbital magnetic fields can also be taken into account in 
our formalism but give no
qualitative changes.  In Eq.~(\ref{hd}), 
the $x (y)$ component of the operator
${\bf a}=(a_x,a_y)$ acts in spin space,
\begin{equation}\label{vecpot}
{\bf a}  =  \alpha \left( \begin{array}{cc} 
\sin\theta \ \sigma_x - \cos\theta \ \sigma_y \,, &
\cos\theta \ \sigma_x - \sin\theta \ \sigma_y \end{array}\right),
\end{equation}
and contains the combined effect of Rashba ($\alpha_R$) and linear
Dresselhaus ($\alpha_D$) couplings via 
\begin{equation}\label{coup}
\alpha=\sqrt{\alpha_R^2+\alpha_D^2},\quad 
\sin \theta=\alpha_D/\alpha.
\end{equation}  
These two are generally the most important
SO couplings in QDs based on 2DEG geometries.  
The superconducting banks are described as
3D $s$-wave BCS models. 
The Hamiltonians of the left and right superconducting electrodes 
have the standard BCS form,
\begin{equation}
\label{BCS}
H_{j=L/R} = \sum_{{\bf k}, \, \sigma = \uparrow,\downarrow}
\xi_{\bf k} \, \Psi^\dagger_{j {\bf k} \sigma} \Psi_{j {\bf k} \sigma} +
\sum_{\bf k} \left( \Delta \, e^{\mp i \phi/2} \,
\Psi^\dagger_{j {\bf k} \uparrow} \Psi^\dagger_{j (-{\bf k}) \downarrow}
+ {\rm h. c.} \right),
\end{equation}
with $\xi_{\bf k} = k^2/(2m) - \mu$ and 
electron creation operators $\Psi^\dagger_{L/R,{\bf k},\sigma}$, 
and carry the relative phase  $\phi$.
For simplicity, we assume the same BCS gap $\Delta$ on both banks;
in experiments, both contacts are typically created by the same 
lithographic process.
Finally, $H_T$ describes tunneling between the leads and the QD,
\begin{equation}\label{tunnel}
H_T=\sum_{L/R,{\bf k},n}  t_{L/R,n} 
\Psi_{L/R,{\bf k}}^\dagger  U_{L/R} \ d_{n} + {\rm h.c.} ,
\end{equation}
where for now $U_{L/R}=1$ (but see below), and
the index $n$ labels the QD eigenstates for $\alpha={\bf b}=0$, 
\begin{equation} \label{conf}
 \left( -\frac{1}{2m}\nabla^2 + V({\bf r}) \right) \chi_n({\bf r}) =
\epsilon_n \chi_n({\bf r}) \,,
\end{equation}
with $d_\sigma({\bf r}) =\sum_n \chi_n({\bf r}) d_{n\sigma}$.
To simplify our expressions, we make the inessential assumption of 
${\bf k}$- and spin-independent real-valued
tunnel amplitudes $t_{L/R,n}$ between the $L/R$ banks and the QD state
$\chi_n$. 
Although the connection between the superconducting leads 
and the dot appears in the form of a tunneling Hamiltonian, 
we treat these terms below to all orders
by integration over the lead degrees of freedom. 
This provides justification for ignoring Coulomb interactions on the QD.
Over the last decade, several works \cite{flensberg_matveev_nazarov}
have shown that in the limit of high transparency, i.e. 
for tunneling rates large compared to the dot charging energy, 
charge quantization (Coulomb blockade) effects are strongly 
suppressed due to large charge fluctuations.
Note also that an observable Josephson current usually requires 
high-transparency contacts.

Without loss of generality, we now choose a coordinate system
where the tunnel contacts are located at $(x,y)=(\mp L/2,0)$. 
It is then convenient to eliminate the $a_x$-term by 
a unitary transformation on the dot fermions \cite{sun},
\begin{equation}\label{unitar}
d({\bf r}) \to e^{-ix a_x}  d({\bf r}).
\end{equation}
When expanded into the basis (\ref{conf}),
the transformed $H_D$ is 
\begin{equation}\label{hdt}
H_D = \sum_{n} \left(\epsilon_n 
- \frac{\alpha^2}{2 m}\right) d^\dagger_{n} d^{}_{n}
+ \sum_{nn'} d_n^\dagger \left( {\bf A}_{nn'} + {\bf B}_{nn'} \right)
\cdot \, \vec\sigma \, d_{n'} ,
\end{equation}
where the transformed $a_y$ is encoded in
\begin{eqnarray} \label{afield}
{\bf A}_{nn'} = -i\frac{\alpha}{m} \int d{\bf r} 
 \chi_n^\ast({\bf r}) [ \partial_y\chi_{n'}({\bf r})]  
\left( \begin{array}{c}
\cos\theta [1-2\cos(2\theta)\sin^2(\alpha x)] \\
-\sin\theta [1+2\cos(2\theta)\sin^2(\alpha x)] \\
-\cos(2\theta) \sin(2\alpha x) \end{array} \right),
\end{eqnarray}
and the Zeeman field leads to 
\begin{equation}\label{bfield}
{\bf B}_{nn'}\cdot \vec\sigma =  \int d{\bf r} \,
\chi_n^\ast ({\bf r})  \left( {\bf b} \cdot e^{ixa_x} \vec\sigma 
e^{-ix a_x} \right) \chi_{n'}({\bf r}).
\end{equation}
Since the dot electron wave function $\chi_n({\bf r})$ in Eq.~(\ref{conf}) 
can always be chosen real-valued, the 
$A_{i,nn'}$ ($ B_{i,nn'}$) with $i=x,y,z$ represent
 antisymmetric (symmetric) Hermitian matrices in the QD level space. 
Finally, the transformation (\ref{unitar}) yields the $2\times 2$
spin matrices
\begin{equation}\label{url}
U_{L/R} = \left( \begin{array}{cc}
\cos(\alpha L/2) & \mp \sin(\alpha L/2) e^{-i\theta} \\
\pm \sin(\alpha L/2) e^{i\theta} & \cos(\alpha L/2)
\end{array}\right)
\end{equation}
entering $H_T$ in Eq.~(\ref{tunnel}). 
The oscillatory dependence on the length $L$ between the tunnel
contacts appearing for $\alpha \neq 0$ is 
due to a spin precession phase \cite{datta}.

\section{Josephson current}
\label{Sec.III}

The equilibrium Josephson current at temperature $T$ follows from the 
free energy $-T \ln Z$ via 
\begin{equation}\label{josephs}
I(\phi)= -\frac{2e}{\hbar} T \partial_\phi \ln Z(\phi),
\end{equation} 
where the superconducting phase difference enters the
 BCS Hamiltonian (\ref{BCS}).
We adopt a functional-integral representation of the partition
function $Z$, which here requires a slightly non-standard formulation due to 
the spin-flip terms.
Given the fact that the Hamiltonian is quadratic in the fermion
operators, the calculation should proceed in a straightforward manner.
In the absence of spin flips, such quadratic Hamiltonians are typically 
dealt with by introducing Nambu spinors.
For the present case (see Appendix), 
the presence of spin flip processes makes it necessary to introduce two types 
of Nambu spinors to describe the dot. 
In the end, the effective action describing the quantum dot 
coupled to the leads can nevertheless be written in terms of  
Grassmann fields $d_{n\sigma}(\tau)$
and $\bar d_{n\sigma}(\tau)$ for the dot fermions ($\tau$
is imaginary time). For a multilevel dot, we thus form the Nambu spinor
for dot level $n$,
\begin{equation} \label{fourier}
\left ( \begin{array}{c} d_{n\uparrow}(\tau)\\
 {\bar d}_{n\downarrow}(\tau) \end{array}\right) =
 \sqrt{T}\sum_\omega e^{-i\omega \tau} D_{n}(\omega) ,
\end{equation}
with fermionic Matsubara frequencies $\omega=(2l+1)\pi T$ (integer $l$).
Similarly, we define the Nambu multispinor
\begin{equation}
\label{multispinor}
D(\omega)=(D_1(\omega),D_2(\omega),\ldots)^T,
\end{equation}  
and then integrate out the lead fermions. 
As a result of this operation, $Z=\int {\cal D}
(\bar D, D) e^{-S}$
with the action  
\begin{equation}\label{act}
S = T \sum_{\omega>0} \left( D(\omega), \bar D(-\omega) \right) 
\hat S_\omega \left( \begin{array}{c} D(\omega) \\ \bar D(-\omega) 
\end{array}\right) .
\end{equation} 
The apparent doubling of spinor space reflects our above discussion.
This doubling does not create a double-counting problem, since only
positive Matsubara frequencies $\omega$ appear in Eq.~(\ref{act}).
After straightforward but lengthy
algebra, with $A_{\pm,nn'}=A_{x,nn'}\pm iA_{y,nn'}$
(and likewise for ${\bf B}$), see Eqs.~(\ref{afield}) and (\ref{bfield}),
we find
\begin{equation} \label{sw}
\hat S_\omega = \left( \begin{array}{cc} 
G^{-1}_\omega+ A_{z} \sigma_z + B_z & A_-\sigma_x+i B_-\sigma_y \\
A_+\sigma_x- i B_+ \sigma_y & - [G^{-1}_{-\omega}]^T + A_z \sigma_z - B_z
\end{array}\right).
\end{equation}
The $2\times 2$ structure refers
to this doubled space, where each of the four entries represents
a matrix in multispinor Nambu space, i.e. the
Pauli matrices  in Eq.~(\ref{sw}) act in the Nambu (spin) sector,
while ${\bf A}$ and ${\bf B}$ operate on the level space.
Spin-flip transitions caused by SO processes are described 
by the off-diagonal entries $A_\pm$ and $B_\pm$. 
Finally, $G_\omega$ denotes the QD Green's function after
integration over the leads but in the absence of SO couplings. 
This Green's function is derived and specified in the
Appendix using the  wide-band approximation, see Eq.~(\ref{gdef}),
with the leads characterized by a constant normal-state 
density of states $\nu$ 
(assumed identical in both banks), and with
the hybridization matrix 
\begin{equation}
\label{hybri} 
\Gamma_{L/R,nn'}=\pi\nu | t_{L/R,n} t_{L/R,n'}|.
\end{equation}
Note that eventually the $U_{L/R}$ factors do not contribute to the action, 
see Eq.~(\ref{sw}), and thus the precession phase $\alpha L$
can appear only via ${\bf A}$ and ${\bf B}$.
We finally obtain from Eq.~(\ref{josephs}) the 
equilibrium Josephson current-phase relation,
\begin{equation} \label{final}
I(\phi) = -\frac{2e}{\hbar} T \sum_{\omega>0} 
\partial_\phi \ln \det \hat S_\omega.
\end{equation}
We stress that this result (given the model) is exact and does 
not involve approximations.
The calculation of the Josephson current requires only a simple
numerical routine and allows for a quantitative comparison to 
experimental data.  

\section{Applications} \label{Sec.IV}

In order to get insight into the relevant physics, 
we here analyze the simplest
cases of a single and of two spin-degenerate QD levels with
symmetric hybridization matrix, 
$\Gamma_{L,nn'}=\Gamma_{R,nn'} \equiv \Gamma_{nn'}/2$.
While our conclusions are general, some
coefficients below are model-dependent. They are derived for $V({\bf r})$
given as hard-box confinement along the $x$-axis, 
 $-L/2\leq x\leq L/2$, plus a harmonic
transverse confinement of frequency scale $\omega_\perp$. 
The level index $n=(n_x,n_y)$ then contains the respective integer 
quantum numbers
$n_x\geq 1$ and $n_y\geq 0$, with eigenenergy (up to an additive constant 
related to a gate voltage)
\begin{equation}\label{epsn} 
\epsilon_n= \frac{(\pi n_x/L)^2}{2m} +\omega_\perp(n_y+1/2)
\end{equation}
in Eq.~(\ref{conf}). Note that
Eq.~(\ref{final}) contains not only the 
contributions of Andreev bound states but represents the
full Josephson current.  The necessity of computing the full
current for finite $L$ has been stressed recently \cite{kamenev}.

\subsection{Single dot level}

\begin{figure}
\scalebox{0.35}{\includegraphics{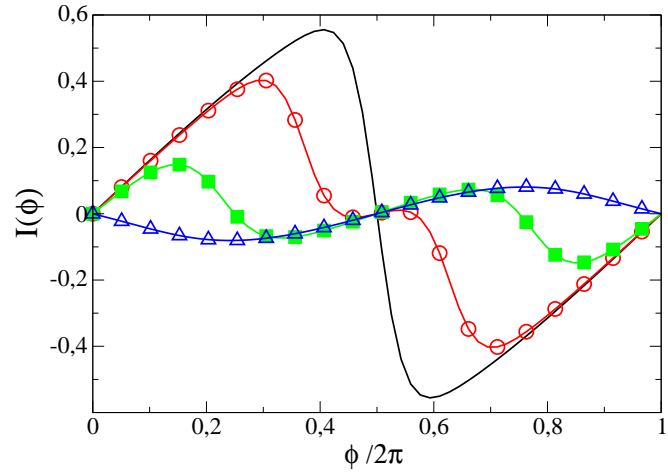}} 
\caption{ \label{fig1} (Color online) 
Josephson current in units of $e\Delta/\hbar$
through a single-level QD with $\epsilon = 0, T=0, \Gamma = 2$
for $B = 0$ (solid curve), $0.4 \Gamma$ (circles), $0.8 \Gamma$ (squares), 
and $1.2 \Gamma$ (triangles).}
\end{figure}

Let us first address a {\sl single QD level},
where we reproduce and extend previous results 
\cite{chalmers,krive}.  
The antisymmetry of ${\bf A}_{nn'}$ implies ${\bf A}=0$,
and thus SO couplings do not affect the Josephson current
in the absence of a Zeeman field \cite{chalmers}.  With 
$B=|{\bf B}|$ given by 
\begin{equation}\label{b2}
B^2 =  (b_x\sin \theta-b_y \cos\theta)^2  + F_{n_x}
\left(b_z^2+(b_x\cos\theta+b_y\sin\theta)^2\right),
\end{equation}
where $F_n$ is defined as
\begin{equation}\label{Fal}
F_n(\alpha L) = \left(\frac{\sin(\alpha L)}{\alpha L(1- (\alpha L/\pi n)^2)}
\right)^2  ,
\end{equation}
the determinant entering Eq.~(\ref{final}) is
\begin{equation}\label{detss}
{\rm det} \hat S_\omega = 
4\gamma_\omega^2 B^2 +
\left (\epsilon^2 + \gamma^2_\omega + \frac{\Gamma^2\Delta^2\cos^2(\phi/2)}{
\omega^2+\Delta^2} - B^2\right)^2,
\end{equation}
where $\gamma_\omega= \omega(1+\Gamma/\sqrt{\omega^2+\Delta^2})$.
Here we consider a situation where transport proceeds through a single 
resonant level with longitudinal quantum number 
$n_x\geq 1$, i.e. $\epsilon\equiv \epsilon_{n_x}-\alpha^2/2m$ is close to zero,
 while the dot stays in the transverse ground state $n_y=0$.
Under the condition $\omega_\perp \gg \delta \epsilon \gg |{\bf b}|$ for the
level spacing $\delta \epsilon$, i.e. for a
quasi-1D situation, all states besides the $n_x$ state close
to resonance can be neglected, and one effectively has a single-level
dot.  Clearly, SO effects now
can only enter via the effective magnetic field (\ref{b2}).
Compared to the Andreev level spectrum of the $\alpha=b=0$ QD connected to 
superconducting electrodes, the combination of SO and Zeeman 
couplings gives rise to a splitting of the Andreev levels,
giving in total four distinct levels. Each level carries a definite 
spin with both spin up/down levels above/below the chemical potential.   
Superficially, the structure of this Andreev spectrum is somewhat analogous 
to the one of Ref.~\cite{chtchelkatchev}. However, it is actually 
different because in Ref. \cite{chtchelkatchev} the levels refer to zero spin, 
spin up, spin down, and a spin singlet state, respectively. 
With Eq.~(\ref{detss}) the Josephson current
appears in precisely the form of the mean-field solution of
the interacting Anderson dot problem obtained 
by Vecino {\sl et al.} \cite{vecino}.
This implies that their results can be taken over (with their
exchange energy $E_{ex}$ being our $B$).
In particular, the phase diagram in the $B$-$\epsilon$ plane
exhibits rich behavior with four different phases
including $\pi$-junction behavior, see Fig.~3 of Ref.~\cite{vecino}.
In Fig.~\ref{fig1}, the Josephson current is plotted as function of the 
phase difference. For simplicity, we consider $\epsilon=0$
and gradually increase the effective magnetic field $B$ given in 
Eq.~(\ref{b2}). The transition to a $\pi$ junction with negative
critical current occurs when $B>\sqrt{\Gamma^2+\epsilon^2}$. 

\begin{figure}
\scalebox{0.35}{\includegraphics{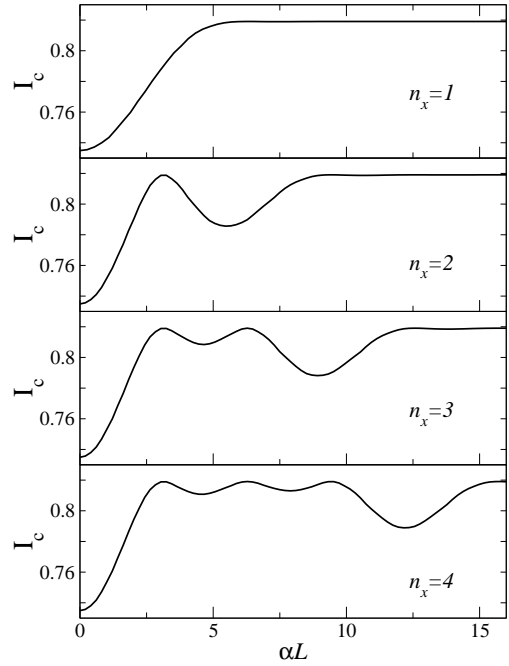}} 
\caption{ \label{fig2} SO-induced oscillations of the 
critical current $I_c$ (in units of $e\Delta/\hbar$) in a single-level dot
as a function of $\alpha L$ for ${\bf b}=(0,0,0.2\Gamma)^T, \Gamma=10\Delta,
T=0.05\Delta$.  The dot is taken in the transverse ground state $n_y = 0$,
with only one resonant level $\epsilon_{n_x}=0$,
for various $n_x$. }
\end{figure}

For $b_x=b_y=0$, the oscillatory behavior of $B$ due to the dependence
of $F_n$ on the spin precession phase $\alpha L$ is most pronounced, 
with $B=0$ for $\alpha L=2\pi l$ (integer $l$).
This oscillation may then persist in the Josephson current, suggesting
the appearance of Datta-Das like oscillations in the critical
current. These oscillations are displayed in Fig.~\ref{fig2};
for $n_x=1$, such oscillations are not yet observable,
but they become visible for $n_x>1$. 
Note that for convenience, the plots in Fig.~\ref{fig2} ignore the 
$\alpha^2/(2m)$ shift of $\epsilon_{n_x}$, as is appropriate 
for small $\alpha$;
alternatively, one could adjust a gate voltage on the quantum dot
in order to bring the level into resonance. 
For $\Gamma \gg \Delta, |{\bf b}|$, 
the amplitude of the current oscillations is of the order of 
$(e \Delta/ \hbar) (|{\bf b}|/\Gamma)^2$.
The oscillation period in the critical current 
is roughly set by $\alpha L=\pi$, similar to 
the normal-state case of the Datta-Das transistor \cite{datta}.  
Moreover, by systematic variation of the magnetic field (${\bf b}$)
direction within the $x$-$y$ plane, one could measure
the SO angle $\theta$ from the Josephson current, see Eq.~(\ref{b2}).

\subsection{Two-level dot}

In order to get SO-related effects on the Josephson current without
a magnetic field, ${\bf b}=0$, 
 we study a QD with {\sl two levels} as a minimal model.
As a concrete example, consider the dot states given by the first two 
oscillator eigenstates ($n_y=0,1$) in the longitudinal ground
state ($n_x=1$), such that the energy levels are encoded in
$\epsilon_2-\epsilon_1=\omega_\perp$
and $\epsilon_0=(\epsilon_1+\epsilon_2)/2$.
[Note that $n_y\neq n_y'$ is necessary 
for ${\bf A}_{nn'}\neq 0$, see Eq.~(\ref{afield}).]
Now all matrix elements of  ${\bf B}$ and ${\bf A}$ are zero except
for ${\bf A}_{12}= -{\bf A}_{21}$, 
with $A\equiv |{\bf A}_{12}|$ given by
\begin{equation}\label{ael} 
A^2=  \frac{\alpha^2\omega_\perp}{2m} 
\left( \sin^2(2\theta)+ F_1(\alpha L) \cos^2(2\theta)  \right).
\end{equation}
One easily checks that ${\rm det} \hat S_\omega$ and therefore
the Josephson current now depends on $\alpha$
and $\theta$ exclusively through Eq.~(\ref{ael}).

\begin{figure}
\scalebox{0.71}{\includegraphics{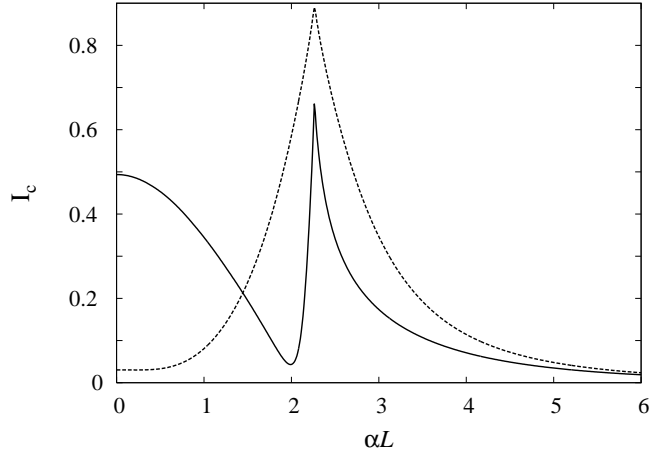}} 
\caption{ \label{fig3} Zero-field critical current
as a function of $\alpha$ in a two-level quantum dot. 
We take $\epsilon_0=0$, $\omega_\perp=20$~meV, $\theta=0$, $\Delta=1$~meV, $L=20$~nm, $m = 0.035 m_e$, and $T=0.01 \Delta$. 
(These values are appropriate for Nb contacts.)
Results are shown for two cases  
(where $\Gamma_{12}=\sqrt{\Gamma_{11}
\Gamma_{22}}$), namely either $\Gamma_{11}=20\Delta$ and $\Gamma_{22}=0$
(only one level couples to the leads, solid line)  or 
 $\Gamma_{22}=\Gamma_{11}=10\Delta$ ('democratic tunneling', dashed line).
}
\end{figure}

Figure \ref{fig3} shows typical numerical results based on Eq.~(\ref{final}) 
for the critical current $I_c$ as a function of $\alpha$
for two choices of the hybridization parameters $\Gamma_{nn}$,
with $\Gamma_{12}= \sqrt{\Gamma_{11}\Gamma_{22}}$.  Here we
consider a pure Rashba coupling ($\theta=0$),  and
the dot levels 
are determined by $\epsilon_0=0$ and $\hbar\omega_\perp=20$~meV.
The shown range for $\alpha L$ can be
realized in InAs-based devices \cite{nitta}, and thus
the supercurrent can be strongly modified by experimentally
relevant SO couplings in multi-level quantum dots,
with pronounced minima or maxima in $I_c$.
The apparent cusp-like maximum in Fig.~\ref{fig3} is
smooth and does not represent singular behavior.

Note that the results displayed in Fig.~\ref{fig3} are for 
large $\Gamma/\Delta$, 
where the Josephson current is predominantly carried
by the Andreev bound state contribution $I_A(\phi)$. 
The latter can be analytically evaluated from 
Eqs.~(\ref{sw}) and (\ref{final}) for 
$(\Gamma_{11}+\Gamma_{22})/\Delta\gg 1$, and (at $T=0$) is expressed in
terms of an effective transmission probability ${\cal T}_0$ 
as in Refs.~\cite{jj,feigelman}, 
\begin{equation}\label{andreev}
I_A(\phi) = \frac{e\Delta}{2\hbar} \frac{{\cal T}_0 \sin\phi}
{\sqrt{1-{\cal T}_0\sin^2(\phi/2)}},
\end{equation}
where we find ${\cal T}_0=1/(1+z^2)$ with
\begin{equation}
z= \frac{ A^2+\omega_\perp^2/4-\left( \epsilon_0-\alpha^2/2m \right)^2} 
{\left( \Gamma_{11}+\Gamma_{22} \right) \left( \epsilon_0-\alpha^2/2 m \right) +
\left(\Gamma_{11}-\Gamma_{22} \right)
\omega_\perp/2 }. 
\end{equation}
The minimum in $I_c$ is thus expected when ${\cal T}_0\simeq 0$,
in accordance with the values seen in Fig.~\ref{fig3}.
(For the dashed curve in Fig.~\ref{fig3}, 
the $I_c$ minimum is at $\alpha=0$.) 
On the other hand, the maximum in $I_c$ (where $I_c$ may 
exceed the $\alpha=0$ value) is found for ${\cal T}_0=1$,
again explaining the numerical result in Fig.~\ref{fig3}.
Here the level shift $\alpha^2/(2m)$ brings one dot level into resonance
and thereby causes the enhanced supercurrent.
This simple reasoning also explains why
the position of the $I_c$ maximum depends only very weakly 
on the $\Gamma_{nn'}$. 

The resonance condition, however, depends trigonometrically 
on the SO coupling $\alpha$ via $A^2$, see Eq.~(\ref{ael}). 
It results from the interplay between SO couplings and 
the multilevel nature of the QD.
By the description in terms of Andreev bound states, we can also 
understand why for large dots (long or wide), 
the effect of the SO coupling is negligible.  
First, for a `long' dot, taking $\theta=0$,
$A^2$ decreases as $L^{-2}$, and therefore the SO effects on
the Josephson current vanish as $L\to \infty$.
Second,  when the lateral dimension of the dot is large, 
the confinement frequency $\omega_\perp$ becomes small. Now
$A^2$ contains $\omega_\perp$ as a prefactor, and hence 
is once again suppressed.  In both situations, 
the effect of the SO interaction 
is merely reduced to a shift of  energy levels.

A qualitative argument in favor of such a resonance process
can be reached as follows.  Schematically, when a Cooper pair 
enters the dot from the left, either its electrons occupy 
the same transverse level
or they can choose different `paths' (different 
transverse levels), in a manner quite similar to a cross-Andreev scattering 
process \cite{martin_feinberg}.  The SO interaction acts differently
in these two levels because the longitudinal momenta 
of both states are not identical. 
Two electrons entering the QD from, say, the left 
electrode have initially antiparallel spins, 
but their respective spins now precess at different 
rates because of this mismatch in longitudinal momentum. 
Depending on the value of  $\alpha L$, the two electrons which then exit 
the QD at the right side may, however, be brought back to an antiparallel 
configuration, leading to a resonance in the critical current caused
by the presence of SO couplings.
On the other hand, if the spins do not reach the antiparallel 
configuration, the Josephson current will be reduced. 
 
\begin{figure}
\scalebox{0.71}{\includegraphics{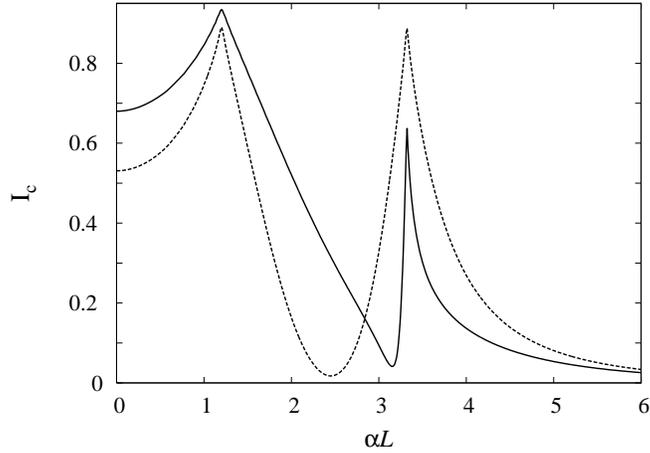}} 
\caption{ \label{fig4} Same as Fig. \ref{fig3}, but for
$\epsilon_0 = 15$~meV.}
\end{figure}

Note that more than one resonance can be achieved
by a careful choice of parameters.
This effect is illustrated in Fig.~\ref{fig4}.
For a multi-level dot with small confinement frequency, 
the multiple resonances associated with all possible pairs of paths
(pairs of levels) will start to overlap significantly.
It is then natural to expect that in the limit of very wide
quantum dots, these averaging effects will wash out any SO-related
structures in the   Josephson current, 
consistent with the results of Ref.~\cite{feigelman} for 
infinite width.   

\section{Conclusion}\label{Sec.V}

To conclude, we have studied the Josephson current through a 
multi-level quantum dot in the presence of Rashba and Dresselhaus
spin orbit couplings. 
In the absence of electron-electron interactions
on the dot, this problem is exactly solvable, a simple consequence of the fact 
that the Hamiltonian is quadratic in the creation/annihilation operators.
(Our model consists of a noninteracting quantum dot including spin
flip processes and connected to BCS superconducting leads.) 
Nevertheless, the combination of superconducting leads
and spin-flip terms renders a calculation of the Josephson 
current a quite technical task, which we chose to address
via functional integral techniques.    
We have explicitly shown results for one and two levels, but our general 
expressions can be applied to arbitrary situations.

For a single dot level, spin-orbit effects cancel out
unless a magnetic Zeeman field is included.
In this case, 
we predict spin precession (Datta-Das) effects 
in the Josephson current, i.e. oscillations as a function of the
effective length of the dot.
These oscillations have  amplitude of the order of a few tenths 
of the nominal critical current, which should be observable.
They result from  the interplay between the  period
of the oscillating effective magnetic field 
(caused by the combined effects of the 
Zeeman and Rashba interactions) with the wavelength of the 
longitudinal modes in the dot.

More interestingly, for the case of a double dot, spin precession 
effects show up even in the absence
of an external magnetic field. This is a novel effect 
in the field of Josephson physics for devices subject to spin-orbit coupling.  
The supercurrent can be drastically modified, either containing
sharp peaks or being largely suppressed.
The experimental observation of such peaks could constitute 
evidence for spin-orbit effects in a superconducting transistor.   

Possible extensions of this work could include Coulomb interaction
effects in the dot, which 
will be important when the tunneling rates become comparable to the 
dot charging energy. While the inclusion of such effects is 
beyond the scope of this paper, the present 
formulation of the problem with functional integral techniques 
can be adapted to include them in an approximate manner.
Nevertheless, we stress again that the high transparency limit 
considered here is in fact quite relevant in the light of recent experiments
where quantum dots are embedded in a Josephson setting
using InAs nanowires \cite{doh,vandam}.
   
\acknowledgments

R.E. and L.D. acknowledge support from the EU network HYSWITCH. 
T.M. acknowledges support from a CNRS 
Action Concert\'ee Nanosciences and from an ANR PNANO Grant.  

\begin{appendix}
\section{}

In this appendix, we provide some details
concerning the derivation of the action in Sec.~\ref{Sec.II}.
The partition function is here calculated by using
the path-integral approach. For this purpose, we
rewrite the Hamiltonian in terms of Grassmann-Nambu spinors.
We introduce two types of Nambu spinors ('spin-up' and 'spin-down') 
for the QD fermions,
\[
d_n = \left( \begin{array}{c} d_{n\uparrow} \\ 
\bar{d}_{n\downarrow} \end{array} \right) ~,~~~
d^c_n = \left( \begin{array}{c} d_{n\downarrow} \\ 
\bar{d}_{n\uparrow} \end{array} \right) ,
\]
while the BCS leads are described by the conventional 'spin-up' Nambu spinor,
$\Psi_{j{\bf k}} = ( \Psi_{j {\bf k} \uparrow},
\bar{\Psi}_{j (-{\bf k}) \downarrow} )^T.$
With the notation introduced in Sec.~\ref{Sec.II},
we can then effectively write
\begin{eqnarray*}
H_D &=& \sum_{nm} \bar{d}_n  \left[ \left( 
\epsilon_n - \frac{\alpha^2}{2 m} \right) \, \delta_{nm} \,
\sigma_z  + A_{z,nm} \, \sigma_z + B_{z,nm} \right] d_m
\\ &+&  \frac12 \sum_{nm}  \left[\ \bar{d}_n \left(
A_{-,nm}  + B_{-,nm} \ \sigma_z \right) d^c_m
+ \bar{d}_n^c \left( A_{+,nm}  + B_{+,nm} \ \sigma_z \right) d_m  \right].
\end{eqnarray*}
In spinor notation, the tunneling Hamiltonian reads
\[
H_T = \sum_{j=L,R} \sum_{{\bf k} n} \left[ \, \bar{\Psi}_{j {\bf k}} \, \left( T_{jn} \, d_n +
{\cal T}_{jn} \, d_n^c \right) + {\rm h.c.} \right].
\]
After gauging out the phase $\phi$ from the leads, 
\begin{eqnarray*}
T_{j = L/R=\pm, n} &=& \cos(\alpha L/2) \, e^{\pm i \sigma_z \phi / 4} \,
\left( \begin{array}{cc} t_{jn} & 0 \\ 0 & - t^\ast_{jn} \end{array} \right)  ,
\\
{\cal T}_{j, n} &=& \mp \sin (\alpha L/2) \, e ^{-i \theta} \,
e^{\pm i \sigma_z \phi / 4} \,
\left( \begin{array}{cc} t_{jn} & 0 \\ 0 & t^\ast_{jn} \end{array} \right) ~.
\end{eqnarray*}
After integrating out the $\Psi_{j{\bf k}}$ and taking into account
the relation between the Fourier transformed
'spin-up' and 'spin-down' Nambu spinors for the QD, see Eq.~(\ref{fourier}),
\[
D_n(\omega)\equiv\frac{1}{\sqrt{T}}\int d\tau e^{i\omega\tau} d_n(\tau) ~,~~~
\sigma_x\bar{D}_n(-\omega)\equiv \frac{1}{\sqrt{T}}\int d\tau e^{i\omega\tau} 
d^c_n(\tau),
\]
we obtain the effective action
\begin{equation} \label{SeffXi}
S  =  S_D - T \sum_\omega \sum_{nm}
\left( \begin{array}{cc}\bar{D}_n (\omega) & D_n (- \omega) \end{array} \right)
\sum_j \left( \begin{array}{cc}
T^\dagger_{jn} \, g_\omega \,  T_{jm} &
T^\dagger_{jn} \, g_\omega \,  {\cal T}_{jm} \sigma_x \\
\sigma_x {\cal T}^\dagger_{jn} \, g_\omega \,  T_{jm} &
\sigma_x {\cal T}^\dagger_{jn} \, g_\omega \,  {\cal T}_{jm} \sigma_x
\end{array} \right)
\left( \begin{array}{c} D_m (\omega) \\ 
\bar{D}_m (- \omega) \end{array} \right) ,
\end{equation}
where $g_\omega$ denotes the Green's function of the uncoupled leads
\[
g_\omega = \frac{\pi \nu}{\sqrt{\omega^2 + \Delta^2}} \,
\left( i \omega + \Delta \, \sigma_x \right).
\]
Notice that there is no contribution 
from the gauged-out spin-flip terms describing spin precession 
along the transport direction, since
in the second term of Eq.~(\ref{SeffXi}), all $d d$ terms vanish.
Finally, the action of the closed QD is
\begin{eqnarray*}
S_D &=& T \sum_\omega \sum_{nm} \left[ \, 
\bar{D}_{n} (\omega)  \left[ 
\left( - i \omega + 
\left( \epsilon_n - \frac{\alpha^2}{2 m} \right) \sigma_z \right) \, \delta_{nm} + 
A_{z,nm} \, \sigma_z + B_{z,nm} \right] D_{m}(\omega)
\right.
\\ 
&& \left.
+ \frac12  \left[\, \bar{D}_{n}(\omega) \left(
A_{-,nm} \, \sigma_x + i B_{-,nm} \, \sigma_y \right) \bar{D}_{m} (-\omega)
+ D_{n} (-\omega) \left(
A_{+,nm} \, \sigma_x -i B_{+,nm} \, \sigma_y \right) D_{m}(\omega)  \right] 
\right].
\end{eqnarray*}
To simplify calculations, we now assume $t_{jn}^\ast = t_{jn}$.
Introducing the multispinor (\ref{multispinor}),
after some algebra, we then find the effective action (\ref{act}) 
with Eq.~(\ref{sw}), 
where $\hat{G}_\omega$ is the full Green's function of the QD for $\alpha=0$
 (except for the shift in energy),
\begin{equation} \label{gdef}
G^{-1}_{\omega,nn'}  
= \left(-i\omega+ \left[ \epsilon_n-\alpha^2/(2m) \right] \sigma_z \right) 
\delta_{nn'} 
+ \frac{(\Gamma_L+\Gamma_R)_{nn'}}{\sqrt{\omega^2+\Delta^2}} 
\left [ -i\omega+\Delta\cos(\phi/2)\sigma_x \right ] 
+ \frac{(\Gamma_L-\Gamma_R)_{nn'}}{\sqrt{\omega^2+\Delta^2}} 
 \Delta\sin(\phi/2) \sigma_y 
\end{equation}
with the hybridization matrix given by Eq.(\ref{hybri}).
\end{appendix}

\newpage

\newpage 

\end{document}